# Comment on the perspective article "Thermodynamic uncertainty relations constrain non-equilibrium fluctuations"

R. Dean Astumian
Dept. of Physics
The University of Maine
Orono, ME 04473

In a recent perspective article[1] Horowitz and Gingrich discuss thermodynamic uncertainty relations that have been derived using "stochastic thermodynamics"[2], a theory based on a hypothesis known as local detailed balance. The authors examined the foundations of this theory in their "Box 1: A brief primer on local detailed balance and stochastic thermodynamics", where a kinetic scheme for transport of particles between two reservoirs is presented. Horowitz and Gingrich arrive at a relationship between the probability to cycle through the states in one order vs. the probability to cycle in the reverse order. This relation bears on the extremely important question of what governs directionality when a system is driven away from equilibrium by contact with multiple reservoirs that are not in equilibrium with one another. In the original version of their paper the relation given for this ratio was obviously wrong and contrary to the second law of thermodynamics. Based on our private communications and on the recent paper authored by my colleagues and myself[3], Horowitz and Gingrich accepted the necessity of correcting[4] the error in their Box 1. Unfortunately, in making the correction, the authors introduced an equally serious, if less transparent, mistake, and continue to base their theory on the thermodynamically impossible idea that transitions are mediated by only one of the two reservoirs. In this comment we illustrate how the principle of microscopic reversibility reveals that the true origin of directional cycling amongst a network of states is kinetic asymmetry. This understanding is important in guiding synthesis of molecular machines and other devices designed to exploit transport or catalysis to drive non-equilibrium processes.

In the amended Box 1 of Ref. 1 following the definition of the rate constants (e.g., $\frac{\alpha}{\gamma} = e^{(\mu_{\text{left}} - E_1)/k_B T}$) the authors write "We recognize the terms in all three exponents as the unitless entropy increase of the reservoir on the transfer of an additional unit of energy (and an additional particle), in agreement with the general local detailed balance statement in equation (1)." This assertion is problematic. The entropy produced in the left reservoir when a particle is removed is $\mu_{\text{left}}/k_B T$ as in the original version, not $(\mu_{\text{left}} - E_1)/k_B T$ as in the amended version. Inclusion of $E_1$ suggests that the entropy change when a particle is removed from a reservoir depends on where the particle goes, which it does not. Equation (1), $\ln\left[\frac{k(x,y)}{k(y,x)}\right] = \sigma(x,y)$, further suggests that the ratio of transition constants is a state function. This is not generally true. The ratios of net transition constants ineluctably involve kinetic asymmetry[5-7], a dependence that is obscured by the incorrect Eq. (1). To see why kinetic asymmetry is essential let us consider a more general version of their Fig. 1 **a** and of the kinetic scheme in their Box 1 of Ref. 1. We focus on the cycle $A \rightleftarrows B \rightleftarrows C \rightleftarrows A$. The ratio between the internal (intra-system) transition constants p and q is

$$\frac{p}{q} = e^{(E_1 - E_2)/k_B T} \tag{1}$$

and the ratios of the transition constants for exchange of particles between the system and the reservoirs mandated by microscopic reversibility are

$$\frac{\alpha}{\gamma} = e^{(\mu_{\text{left}} - E_1)/k_B T}; \quad \frac{\beta}{\delta} = e^{-(\mu_{\text{right}} - E_2)/k_B T}; \quad \frac{\kappa}{\varphi} = e^{(\mu_{\text{right}} - E_1)/k_B T}; \quad \frac{\lambda}{\eta} = e^{-(\mu_{\text{left}} - E_2)/k_B T} \tag{2}$$

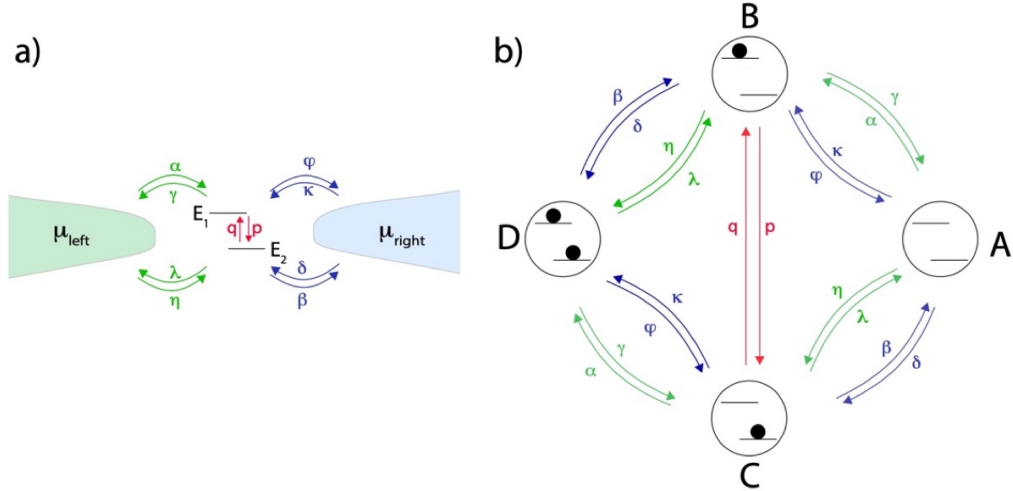

*Figure 1* **a** A model system with two energy levels, each of which can contain 0 or 1 particle, that mediates transport of non-interacting particles between two reservoirs, left (shown in green) and right (shown in blue) held at possibly different chemical potentials $\mu_{\text{left}}$ and $\mu_{\text{right}}$. **b** Kinetic diagram for analyzing the net transport between the two reservoirs and the occupancy of the four possible states of the system, A, B, C, and D. The rate constants into and out of state D reflect the implicit assumption made by Horowitz and Gingrich that the particles do not interact.

In their paper Horowitz and Gingrich implicitly assume that the higher energy ($E_1$) state is inaccessible to the right reservoir, and that the lower energy ($E_2$) state is inaccessible to the left reservoir, and hence they consider only the transition constants $\alpha, \beta, \gamma,$ and $\delta$, treating the transition constants $\kappa, \varphi, \lambda,$ and $\eta$ as being identically zero. This treatment is not thermodynamically consistent - the ratios of forward and backward rate constants are set by thermodynamics. The ratio of the probabilities for clockwise vs. counterclockwise cycling is

$$\frac{\text{prob}(A \xrightarrow{\alpha+\kappa} B \xrightarrow{p} C \xrightarrow{\beta+\lambda} A)}{\text{prob}(A \xrightarrow{\delta+\eta} C \xrightarrow{q} B \xrightarrow{\gamma+\varphi} A)} = \frac{(\alpha+\kappa)}{(\gamma+\varphi)} \times \frac{(\beta+\lambda)}{(\delta+\eta)} \times \frac{p}{q} \tag{3}$$

There are four distinct cycles contained in the expression in Eq. (3). Each cycle is constrained by microscopic reversibility[8,9] based on its the net effect in the environment. For example, the cycle $A \xrightarrow{\alpha} B \xrightarrow{p} C \xrightarrow{\beta} A$ involves transport of a particle from the left reservoir to the right reservoir, and its microscopic reverse $A \xrightarrow{\delta} C \xrightarrow{q} B \xrightarrow{\gamma} A$ involves transport of a particle from the right reservoir to the left reservoir. The ratio of the product of rate constants is given by $k_B T \ln\left(\frac{\alpha p \beta}{\delta q \gamma}\right) = \mu_{\text{left}} - \mu_{\text{right}}$. The expression for a specific cycle was generalized by Hill[10] to

$$\frac{\pi_S}{\pi_{S^\dagger}} = e^{\mathcal{W}_S/k_B T} \tag{4}$$

where $\mathcal{W}_S$ is the net energy exchanged between the system and the reservoirs in the cycle $S$ and where $S^\dagger$ is the microscopic reverse of $S$. The expression in Eq. (4) forms the basis of trajectory thermodynamics[11]. By use of Eq. (2) the ratio in Eq. (3) can be rewritten as

$$\frac{\text{prob}(A \xrightarrow{\alpha+\kappa} B \xrightarrow{p} C \xrightarrow{\beta+\lambda} A)}{\text{prob}(A \xrightarrow{\gamma+\varphi} C \xrightarrow{q} B \xrightarrow{\delta+\eta} A)} = \underbrace{\frac{\left(\frac{\gamma}{\varphi} + e^{(\mu_{\text{right}}-\mu_{\text{left}})/k_B T}\right)}{\left(\frac{\gamma}{\varphi}+1\right)}}_{\mathcal{A}_B} \times \underbrace{\frac{\left(1+\frac{\lambda}{\beta}\right)}{\left(e^{(\mu_{\text{right}}-\mu_{\text{left}})/k_B T}+\frac{\lambda}{\beta}\right)}}_{\mathcal{A}_C^{-1}} \tag{5}$$

where $\mathcal{A}_B$ and $\mathcal{A}_C$ are kinetic asymmetry factors introduced by Astumian and Bier[5] in the context of ATP hydrolysis driven molecular motors. Obviously if $\mu_{\text{right}} = \mu_{\text{left}}$ then $\mathcal{A}_B \mathcal{A}_C^{-1} = 1$, but for $\mu_{\text{right}} \neq \mu_{\text{left}}$ the directionality is controlled by the ratio of "off rate constants". Taking $\mu_{\text{right}} < \mu_{\text{left}}$, for $\frac{\gamma}{\varphi}\frac{\beta}{\lambda} > 1$ then $\mathcal{A}_B \mathcal{A}_C^{-1} > 1$, for $\frac{\gamma}{\varphi}\frac{\beta}{\lambda} < 1$ then $\mathcal{A}_B \mathcal{A}_C^{-1} < 1$, and for $\frac{\gamma}{\varphi}\frac{\beta}{\lambda} = 1$ then $\mathcal{A}_B \mathcal{A}_C^{-1} =$

1. The ratio is independent of $E_1$ and $E_2$. The exact expression $\ln(\mathcal{A}_B \mathcal{A}_C^{-1})$ approaches the expression $\left(\frac{\mu_{\text{left}}}{k_B T} - \frac{\mu_{\text{right}}}{k_B T}\right)$ given by Horowitz and Gingrich in the amended version of their paper[1] only in the limit that $\frac{\gamma}{\varphi} \gg 1$ and that $\frac{\lambda}{\beta} \ll e^{(\mu_{\text{right}} - \mu_{\text{left}})/k_B T}$.

The example[1] in the box 1 is presented as a simple case of the use of local detailed balance and stochastic thermodynamics to show how the transition constants agree with their Eq. (1) for the ratio $\ln\left[\frac{k(x,y)}{k(y,x)}\right] = \sigma(x,y)$, but in fact the transition constants don't agree with this expression. The ratio of the direct transitions between states A and C is

$$\frac{k(C,A)}{k(A,C)} = \frac{\delta + \eta}{\beta + \lambda} = \frac{\left(\beta \, e^{\mu_{\text{right}}/k_B T} + \lambda \, e^{\mu_{\text{left}}/k_B T}\right)}{(\beta + \lambda)} e^{-E_2/k_B T} \qquad (6)$$

which does not reduce to $\sigma(C,A)$ on taking the logarithm. Eq. (6) highlights that in cases where a small system (e.g. an enzyme or a quantum dot, or a synthetic catalytically active molecule) connects two or more reservoirs it is never correct to assume that transitions between states occur by interaction only with one reservoir. Sufficiently far from equilibrium ($\mu_{\text{left}} \gg \mu_{\text{right}}$) we always run into the situation that $\lambda \, e^{\mu_{\text{left}}/k_B T} > \beta \, e^{\mu_{\text{right}}/k_B T}$ even when $\lambda \ll \beta$. By implicitly setting $\eta, \lambda = 0$, which cannot be true since $\frac{\lambda}{\eta} = e^{-(\mu_{\text{left}} - E_2)/k_B T}$, Horowitz and Gingrich obscure this fact and also obscure the origin of directional cycling, which is based solely on kinetic gating[6,12,13].

The example of two spatially separate reservoirs shown in Fig. 1a may seem very far from relating to catalysis driven synthetic[14,15] or biological[8] molecular machines but this is not the case. The substrate and product of a catalyzed reaction form two reservoirs separated, not spatially as in Fig. 1a, but kinetically because the reaction in the absence of the molecular machine would be very slow. A common error involves assigning some transitions to depend on only one reservoir (e.g., binding of ATP, with no binding or release of ADP and Pi) and others to depend on only the other reservoir (e.g., release of ADP and Pi, with no binding or release of ATP) but this approximation is never strictly correct. The importance of kinetic asymmetry for treating systems in contact with several reservoirs was recognized in 1996[5] and is a cornerstone of the development of trajectory thermodynamics in the description of molecular machines[11], but only recently has its general applicability gained widespread acceptance[7,12-15].